\documentclass[aps,pre,preprint,eqsecnum,showpacs,draft,psfig]{revtex4} 
\usepackage{amsmath,bm}
\usepackage{color}

\begin{document}








\title{Vibrational energy transport in molecular wires.}
\author{V.A.Benderskii} 
\affiliation {Institute of Problems of Chemical Physics, RAS, \\ 142432 Moscow
Region, Chernogolovka, Russia} 
\author{A.S. Kotkin} 
\affiliation {Institute of Problems of Chemical Physics, RAS, \\ 142432 Moscow
Region, Chernogolovka, Russia} 
\author{I.V.Rubtsov} 
\affiliation{Department of Chemistry, Tulane University, New Orleans, LA 70118, USA}
\author{E. I. Kats}  
\affiliation{L. D. Landau Institute for Theoretical Physics, RAS, \\ 142432 Moscow
Region, Chernogolovka, Russia}

\begin{abstract}
Motivated by recent experimental observation (see e.g., I.V.Rubtsov, Acc. Chem. Res., ${\bf 42}$, 1385 (2009)) of vibrational energy transport in
$(CH_2O)_N$ and $(CF_2)_N$ molecular chains ($N = 4\, -\, 12$), in this paper we present and solve analytically 
a simple one dimensional model to describe theoretically these data. To mimic multiple conformations of the molecular chains,
our model includes random off-diagonal couplings between neighboring sites. For the sake of simplicity we assume Gaussian
distribution with dispersion $\sigma $ for these coupling matrix elements. Within the model we find that initially locally excited vibrational
state can propagate along the chain. However the propagation is neither ballistic nor diffusion like. The time $T_m$
for the first passage of the excitation along the chain, scales linearly with $N$ in the agreement with the experimental data.
Distribution of the excitation energies over the chain fragments (sites in the model) remains random, and the vibrational
energy, transported to the chain end at $t=T_m$ is dramatically decreased when $\sigma $ is larger than characteristic interlevel
spacing in the chain vibrational spectrum.
We do believe that the problem we have solved is not only of intellectual interest (or to rationalize mentioned
above experimental data) but also of relevance to
design optimal molecular wires providing fast energy transport in various chemical
and biological reactions. 
\end{abstract}

\pacs{03.65, 82.20.B, 05.45.-a, 72.10.-d}

\keywords{quantum dynamics, discrete spectrum, molecular chains, vibrations}

\maketitle

{\bf{Introduction.\, }}Recently \cite{RU1} - \cite{RU5} one of the author of this work (I.R.) together
with his collaborators has observed transport of vibrational energy along $(CH_2O)_N$ and $(CF_2)_N$ molecular chains ($N = 4\, -\, 12$). 
Transfer of the energy from optically excited ''donor'' site is detected by the retardation time $t=T_m$, when the excitation for the
first time arrives to the ''acceptor'' fragment attached to the end of the chain. 
Experimental data show that $T_m$ scales linearly with $N$, and the maximum acceptor site $n=N$ population decays with
a characteristic time scale $\geq 10 \, ps$, comparable with $T_m$.
Our main motivation in this work is to present and to solve
a simple one dimensional model to describe these experimental data. Within the model we compute $T_m(N)$, and excitation energy distribution
(or site $n$ population $\langle |a_n|^2\rangle $).

From the first sight the very existence of such directed energy transport can be 
easily qualitatively understood. Indeed couplings between neighboring chain fragments
lead to delocalized vibrational modes \cite{IY03}. Donor site excitation energy transfers to these modes, and the latter ones 
transport the energy to the acceptor site. However this mechanism for quasi-ballistic energy transport contradicts to common wisdom
based on the radiationless transition theory (see e.g., \cite{UM91}), which predicts very fast equipartition of the excitation energy
over all $N$ sites of the chain. Moreover in multi-atomic fragments (like $CH_2O$ and $CF_2$ investigated in \cite{RU1} - \cite{RU5})
there are several local modes per each site with non-identical coupling matrix elements with the local modes at neighboring sites.
This phenomenon yields to dense quasi-random spectrum of vibrations (similarly to energy levels in large complex nuclei
\cite{PW07}). In this case one can expect mode localization \cite{LG88}, \cite{JA01}, or diffusion like behavior \cite{SN03}, \cite{SH09} 
phenomena rather than ballistic-like plane wave propagation.

Another way to treat this problem is to study directly quantum dynamics of one-particle vibrational excitations. For a uniform
chain with one impurity site in its center, the problem was investigated recently \cite{BK11}, \cite{BK13}, \cite{BK13a}.
The results of these papers suggest that in certain conditions (one delocalized mode weakly coupled to other chain degrees of freedom)
one can expect  realization of the regimes for vibrational excitation propagation observed in the experiments \cite{RU1} - \cite{RU4}.
Motivated by this expectation in what follows we study quantum dynamics of a vibrational excitation created at the donor impurity
site $d$ which is connected by a uniform chain of $N$ identical fragments to the acceptor site $a$ (where the excitation is detected).

{\bf{Model and technical details of its solution.\, }}
For the sake of simplicity we use the units, where the uniform chain site energy $E_n =0$ and nearest neighboring site coupling constant
$C_n =1$. In these units for the donor and acceptor sites we get ($E_d\, ,\, C_d$) and ($E_a\, ,\, C_a$) respectively.
These quantities are defined as the Hamiltonian ${\hat H}$ matrix elements calculated with wave functions in site representation.
They read in self-evident notations as
\begin{eqnarray}
\langle \phi _n |{\hat H}|\phi _n \rangle = 0\, ;\, \langle \phi _n |{\hat H}|\phi _n^\prime  \rangle = \delta _{n^\prime \, ,\, n+1} f_n 
\label{b1}
\end{eqnarray}
for a regular chain $1 \leq n \leq N$ connecting the donor and acceptor sites, and for the latter sites
\begin{eqnarray}
\langle \phi _d |{\hat H}|\phi _d \rangle = E_d\, ,\, \langle \phi _a |{\hat H}|\phi _a  \rangle = E_a \, ;\, 
\langle \phi _d |{\hat H}|\phi _1 \rangle = C_d\, ,\, \langle \phi _N |{\hat H}|\phi _a  \rangle = C_a 
\label{b2}
\end{eqnarray}
Random numbers $f_n$ with $\langle f_n\rangle =1$ and  $\langle (f_n)^2\rangle = \sigma $ is our new ingredient in
this work. There are at least three  sources for these random off-diagonal matrix elements. First for the chains
with multi-atomic fragments studied in \cite{RU1} - \cite{RU5} there are 5 - 7 local modes per each site.
Intersite couplings split the corresponding energy levels into a band with dense spectrum. All 
band levels contribute non-identically into the excitation
energy transfer from the donor to acceptor. Similar roles are played by the conformational degrees of freedom
of the molecular chain and also by surrounding solvent molecules. Although conformational and solvent reorganization
times are much larger than time scales of the order of $10 \, ps$ we are interested in, it makes non-equivalent conditions for
the excitation energy propagation. Thus the matrix Hamiltonian ${\hat H}$ with off-diagonal randomness describes effectively an ensemble
of molecular chains. The secular $(N+2) \times (N+2)$ determinant corresponding to the formulated model has
the following form
\begin{eqnarray}
F(\epsilon)  = 
\left |
    \begin{array}{cccccccc}
        \epsilon - E_d & C_d & 0 & 0 & 0 & 0 & 0 & 0 \\
        C_d & \epsilon & f_1 & 0 & 0 & 0 & 0 & 0 \\
        0 & 0 &  \epsilon & f_n & 0 & 0 & 0 & 0  \\
        0 & 0 &  f_n & \epsilon & f_{n+1} & 0 & 0 & 0  \\
        0 & 0 & 0 & f_{n+1} & \epsilon  & f_{n+2} & 0 & 0  \\
        0 & 0 & 0  & 0 & f_{n+2} & \epsilon  & 0 & 0  \\
        0 & 0 & 0  & 0 & 0 & f_{N-1} & \epsilon  & C_a  \\
        0 & 0 & 0  & 0 & 0 & 0 & C_a & \epsilon - E_a  \\
    \end{array}
\right |
\label{b3}
\, ,
\end{eqnarray}
and the secular equation to find eigen values $\epsilon $ is $F(\epsilon ) =0$.
The Jacobi  form determinant (\ref{b3})  can be written in a more compact form as
\begin{eqnarray}
F(\epsilon ) = (\epsilon - E_d)(\epsilon - E_a) D_N(\epsilon ) - (\epsilon - E_d) C_a^2 D_{N-1}(\epsilon ) 
+ C_a^2 C_d^2D_{N-2}(\epsilon )
\label{b4}
\, ,
\end{eqnarray}
where $D_n(x) = \sum _{k=0}^{[n/2]}(-1)^kA_{n - 2k}(f_k) x^{n-2k}$, $[y]$ is an integer part of $y$, and we introduce the following notations $A_n=1$, $A_{n-2} =\sum _{k=1}^{n-1}f_k^2$, $A_{n-4} = \sum _{k+2 \leq k^\prime \leq n-1}f_k^2 f_{k^\prime }^2$, and so on.

Within the same approach and the Hamiltonian we can expand an arbitrary state time dependent wave function $\Psi (t)$ as
a linear superposition local site wave functions with time dependent amplitudes
\begin{eqnarray}
\Psi (t) = a_d(t) \phi _d + \sum _{n=1}^{N} a_n(t) \phi _n + a_a(t) \phi _a
\label{b5}
\, ,
\end{eqnarray}
and then the Schroedinger equation for $\Psi (t)$  can be formulated as the set of the dynamic equations for the amplitudes
\begin{eqnarray}
i{\dot a}_d = E_d a_d - C_d a_1\, ,\, i{\dot a}_1 = - C_d a_d - a_2\, ,\,i{\dot a}_n = - a_{n-1} +  a_{n+1}\, ,\,i{\dot a}_N = - a_{N-1} - C_a a_a\, ,\,i{\dot a}_a = -C_a a_N + E_a a_a 
\label{b6}
\, ,
\end{eqnarray}
where $\hbar =1$. These equations supplemented by the initial condition $a_d(0)=1$, $a_n(0) =0$, and $a_a(0)=0$ can be solved
by the Laplace transformation, and then after some algebra we end up with the following formally exact solution for a given realization
of random numbers $f_n$
\begin{eqnarray}
a_k(t) = \sum _{k} \exp (i \epsilon _k t) F_k(\epsilon _k \left (\frac{d F}{d \epsilon }\right )^{-1}_{\epsilon = \epsilon _k}
\label{b7}
\, ,
\end{eqnarray}

{\bf{Illustrative Results.\, }}
Expression (\ref{b7}) is our main result and it is ready for further inspection.
In the simplest ideal single chain case, when there is no any randomness, $\sigma =0$, the maximum of the amplitude at the site
$n=N$ determines the first passage time $T_m$. Using Exp. (\ref{b7}) and found in \cite{BK13} representation for the amplitude
\begin{eqnarray}
|a_n(t)| = \frac{1}{2}\left (J_{n-1}(t) + J_{n+1}(t)\right )\, ;\, 2 \leq n \leq N
\label{b8}
\, ,
\end{eqnarray}
valid for the time $t < T_m$ ($J_n$ is the 1-st order Bessel function), we find that the model quantum dynamics 
describes the wave packet motion with an approximately constant (independent of $N$) group velocity $v_g \simeq 2$ (in our units).
The wave packet formed by the amplitudes (\ref{b8}) has very sharp front and weak oscillating tail behind the front.

For $\sigma \neq 0$ we have to deal with an ensemble of chain realizations. Assuming Gaussian random distribution of
the matrix elements $f_n$, we have to average the solution (\ref{b7}). The results, obtained numerically
by using standard Matlab software are presented in the figures 1 - 3. By visual inspection and numeric fitting of the
plots we conclude:
\begin{itemize}
\item
Similarly to the ideal non-random chain, the site $n=N$ population achieves its maximum after the first
passage time $T_m$. As a function of $N$ we can fit $T_m$ as a linear function (see Fig. 1)
\begin{eqnarray}
T_m(N) = T_m^0 + \frac{N}{2}
\label{b9}
\, ,
\end{eqnarray}
where $T_m^0 \propto C_d$ is determined by the rate of the excitation transfer from the donor site to the chain.
\item
There are site population oscillations at $ t> T_m$ related to the waves reflected from the acceptor site.
These oscillations are strongly suppressed when the randomness $\sigma $ increases.
\item
Efficiency of the energy transfer (characterized by the maximum acceptor site population)
decreases with $N$ and $\sigma $ increase. There are two sources for this decay (see Fig. 2).
First source is a natural wave packet spreading (taking place even in a uniform ideal chain \cite{BK13}) has nothing to do with the randomness.
Second source, we are mainly interested in this work, is related to the random fluctuations of the off-diagonal
matrix elements. It reduces dramatically the donor-acceptor excitation energy transfer, when the dispersion $\sigma $
of random fluctuations becomes comparable to the interlevel spacing $\Delta _0 \simeq 2\pi /(N+1)$ in the center of the
uniform chain vibrational band. For $\sigma /\Delta _0 <1$ the sites behind the propagating wave packet remain non-excited
(Fig. 3a), whereas in the opposite case $\sigma /\Delta _0 >1$ there occurs almost equipartition of the excitation energy over all
sites (Fig. 3b). By the numeric fitting of the computed amplitudes presented in the figures 2 and 3, we find
\begin{eqnarray}
\frac{\langle |a_N(\sigma , T_m)|^2\rangle }{|a_N(0, T_m|^2} = \exp (-\sigma ^2(N+1))
\label{b10}
\, ,
\end{eqnarray} 
i.e., the characteristic excitation transfer distance $N_c$ decays proportional to $\sigma ^{-2}$.
\item
It is worth to note, that the critical randomness $\sigma _c \propto 1/{\sqrt N}$ which essentially suppresses the energy transfer
in our model, is much smaller than the corresponding critical value for the localization $\propto N^0$ (\cite{LG88} - \cite{SH09}).

\end{itemize}

{\bf{Conclusions and Perspectives.\, }}
In summary, in this work we propose a simple model explaining semi-quantitatively recent
experimental data \cite{RU1} - \cite{RU5}. The model predicts the excitation first passage from donor to acceptor time
scales linearly with the chain length in a rather broad range of the randomness ($\sigma $). The fact can be understood because
the range of the model parameters where crossover from ballistic like to diffusion like propagation occurs, is much more narrow
than the vibrational spectrum band width. From the experimental data we can also estimate the model parameters. For $N=11$, $CF_2$
chain, we get $v_g \simeq 0.45 \, ps/per\, fragment $, $N_c \simeq 5.5 $, and $\sigma \simeq 0.25$.
Combining these estimations with quantum chemistry ab-initio computations \cite{CH14}, suggesting for the $CF_2$ chain the band of the
delocalized states in the range of $ 900 - 1400 \, cm^{-1}$, $E_d \simeq 2100\, cm^{-1}$, $E_a \simeq 1800\, cm^{-1}$, $C_d \simeq 0.3$,
$C_a \simeq 0.5$, we can be confident that all needed assumptions behind the model are satisfied.
Finally thinking about excitation energy transport in molecular wires, our results suggest that the rigidity of the chain
(i.e., small $\sigma $) is the key requirement for efficient energy transfer. That is why the most favorable conditions
for long distance energy transport can be expected in rod-like rigid systems like DNA or peptide molecules.
We report here only a very brief summary of our results and leave the systematic analysis and applications for future work.

{\bf{Acknowledgments.\, }} E.K. acknowledges the support of RFBR grant No 13-02-00120, and hospitality of the Issac Newton Institute
for Mathematical Science.


\vspace{1cm}

\centerline{Figure Captions}

Figure 1.

Time dependent populations $\langle |a_N|^2\rangle (t)$. (a): $N=10$, and $\sigma $ values for the curves from 1 to 5 are: 0 ; 0.11 ; 0.29 ; 0.43 ; 0.57. (b): $N=2$, and $\sigma $ values for the curves from 1 to 5 are: 0 ; 0.06 ; 0.15 ; 0.22 ; 0.30. $T_m$ are shown by arrows.
The insertion shows the first passage time $T_m(N)$.
The ensemble mean values 
are computed by averaging over 100 realizations.

Figure 2.

Energy transfer efficiency characterized by the ratio $\langle |a_N(t=T_m , \sigma )|^2\rangle /\langle |a_N(t=T_m , \sigma = 0)|^2\rangle $.
The curves from 1 to 4 in the upper plot correspond to $N = 5\, ,\, 10\, ,\, 20\, ,\, 30$. The exponential 
$\exp (-\sigma ^2/\sigma _c(N))$ fitting is shown in the lower plot.

Figure 3.

Time dependent site population $\langle |a_n(t)|^2\rangle $. $N=20$ and discrete time moments are: $t = 1.5\, ;\, 4.0\, ;\, 7.0\, ;\, 10$.
(a): $\sigma = 0.06 $; (b): $\sigma = 0.3 $.

\end{document}